\documentclass[11pt]{article}
\usepackage{graphicx}
\usepackage{epsfig}

\newcommand{\dfrac}[2]{\frac{\displaystyle #1}{\displaystyle #2}}

\newcommand{\ij}{{_i^j}}
\newcommand{\bn}[1]{b^n_{{#1}}}
\newcommand{\sn}[1]{s^n_{{#1}}}
\newcommand{\bw}[1]{b^w_{{#1}}}

\newcommand{\sw}[1]{s^w_{{#1}}}
\newcommand{\fn}[1]{f^n_{{#1}}}
\newcommand{\fo}[1]{f^o_{{#1}}}

\newcommand{\lifetime}[1]{\tau_{{#1}}}
\newcommand{\dm}[1]{\Delta m_{{#1}}}
\newcommand{\ot}[1]{\omega_{{#1}}}
\newcommand{\dmistag}[1]{\Delta \omega_{{#1}}}

\def\bnj#1#2{b^n_{{#1}}{^{#2}}}
\def\snj#1#2{s^n_{{#1}}{^{#2}}}
\def\bwj#1#2{b^w_{{#1}}{^{#2}}}

\def\swj#1#2{s^w_{{#1}}{^{#2}}}
\def\fnj#1#2{f^n_{{#1}}{^{#2}}}
\def\foj#1#2{f^o_{{#1}}{^{#2}}}

\def\beq{\begin{equation}}
\def\eeq{\end{equation}}
\def\bea{\begin{eqnarray}}
\def\eea{\end{eqnarray}}
\def\resp#1{} 

\def\sinphi{\sin(2\beta + \gamma)}

\def\mmiss{m_{\rm miss}}

\def\Dstarp{\Dstar{^+}}
\def\Dstarm{\Dstar{^-}}

\def\btoc{b \rightarrow c \bar u  d}
\def\btou{b \rightarrow u \bar c  d}

\def\btodstpipm{\Bz \rightarrow \Dstarmp\pi^\pm}
\def\btodstpi{\Bz \rightarrow \Dstarp\pi^-}

\def\btodstrhopm{B\rightarrow \Dstarmp\rho^\pm}

\def\dt{\Delta t}
\def\dtErr{\sigma_{\dt}}
\def\dz{\Delta z}

\def\dttrue{\dt_{\rm tr}}

\def\mc{MC}

\def\Brec{B_{\rm rec}}
\def\Btag{B_{\rm tag}}
\def\zrec{z_{\rm rec}}
\def\ztag{z_{\rm tag}}

\def\stag{s_{_{\rm t}}}
\def\smix{s_{_{\rm m}}}

\def\dstpi{{\Dstar\pi}}
\def\dstrho{{\Dstar\rho}}

\def\comb{{\rm comb}}
\def\peak{{\rm peak}}

\def\cont{{q\overline q}}

\def\P{{\cal P}}  
\def\T{{\cal T}}  
\def\M{{\cal M}}  
\def\F{{\cal F}}  
\def\R{{\cal R}}  
\def\A{{\cal A}}  
\def\BG{\hat {\cal G}}
\def\G{{\cal G}}  

\def\r{{r_\dstpi}}
\def\rmeas{r_\dstpi^{\rm meas}}
\def\rp{r'}
\def\rsq{r_\dstpi^{2}}
\def\deltaPhaseP{\delta'}
\def\deltaPhase{\delta_\dstpi}

\def\dir{{\rm dir}}
\def\cas{{\rm cas}}
\def\mis{{\rm miss}}

\def\fdir{f^\dir}
\def\fcas{f^\cas}
\def\fmis{f^\mis}

\def\Tmis{\T^\mis}

\newcommand{\BABARPubYear}    {04}

\newcommand{\BABARConfNumber} {18}
\newcommand{\SLACPubNumber} {10595}
\newcommand{\LANLNumber} {0408038}

\input pubboard/babarsym

\long\def\inst#1{\par\nobreak\kern 4pt\nobreak
    {\it #1}\par\vskip 10pt plus 3pt minus 3pt}

\begin{document}
{\pagestyle{empty}

\begin{flushright}
\babar-CONF-\BABARPubYear/\BABARConfNumber \\
SLAC-PUB-\SLACPubNumber \\
hep-ex/\LANLNumber \\
\end{flushright}

\par\vskip 5cm

\begin{center}
\Large \bf 
Measurement of 
Time-Dependent 
\boldmath \CP Asymmetries and Constraints on $\sin(2\beta+\gamma)$ with
Partial Reconstruction of $\btodstpipm$ Decays 
\end{center}
\bigskip

\begin{center}
\large The \babar\ Collaboration\\
\mbox{ }\\
\today
\end{center}
\bigskip \bigskip

\begin{center}
{\large \bf Abstract}
\end{center}

We present a preliminary 
measurement of the time-dependent \CP-violating asymmetry in
decays of neutral $B$ mesons to the final states $\Dstarmp\pi^\pm$,
using approximately 
$178$ million $\BB$ events recorded by the \babar\ experiment
at the \pep2\ $\epem$ storage ring.  Events containing these decays are
selected with a partial reconstruction technique, in which only the
high-momentum $\pi^\pm$ from the $B$ decay and the low-momentum 
$\pi^\mp$ from the $\Dstarmp$ decay are
used. 
We measure the amplitude of the asymmetry to be  
$-0.041 \pm 0.016~(stat.) \pm  0.010~(syst.)$
and determine bounds on 
$|\sin(2 \beta + \gamma)|$.

\vfill
\begin{center}

Submitted to the 32$^{\rm nd}$ International Conference on High-Energy Physics, ICHEP 04,\\
16 August---22 August 2004, Beijing, China

\end{center}

\vspace{1.0cm}
\begin{center}
{\em Stanford Linear Accelerator Center, Stanford University, 
Stanford, CA 94309} \\ \vspace{0.1cm}\hrule\vspace{0.1cm}
Work supported in part by Department of Energy contract DE-AC03-76SF00515.
\end{center}

\newpage
} 

\begin{center}
\small

The \babar\ Collaboration,
\bigskip

%
B.~Aubert,
R.~Barate,
D.~Boutigny,
F.~Couderc,
J.-M.~Gaillard,
A.~Hicheur,
Y.~Karyotakis,
J.~P.~Lees,
V.~Tisserand,
A.~Zghiche
\inst{Laboratoire de Physique des Particules, F-74941 Annecy-le-Vieux, France }
% ...

\end{center}\newpage


\section{INTRODUCTION}
\label{sec:Introduction}

The Cabibbo-Kobayashi-Maskawa (CKM)
quark-mixing matrix~\cite{ref:km} gives an 
explanation of \CP violation
and is under 
experimental investigation aimed at constraining its parameters. A
crucial part of this program is the measurement of the angle $\gamma =
\arg{\left(- V^{}_{ud} V_{ub}^\ast/ V^{}_{cd} V_{cb}^\ast\right)}$ of
the unitarity triangle related to the CKM matrix.
The decay modes $\Bz \rightarrow {\Dstar}^{\mp} \pi^{\pm}$ have been
proposed for use in measurements of
$\sin(2\beta+\gamma)$~\cite{ref:book}, where $\beta = \arg{\left(-
V^{}_{cd} V_{cb}^\ast/ V^{}_{td} V_{tb}^\ast\right)}$ is well
measured~\cite{ref:sin2b}.
In the Standard Model the decays 
$\Bz \to \Dstarp \pi^-$ and $\Bzb \to \Dstarp \pi^-$
proceed through the $\overline{b} \rightarrow \overline{u}  c  
\overline{d}  $ and
$\btoc$ amplitudes $A_u$ and $A_c$. 
The relative weak phase between the two amplitudes 
in the usual Wolfenstein convention~\cite{ref:wolfen}
is $\gamma$.
When combined with $\Bz \Bzb$ mixing, this yields a weak phase
difference of $2\beta+\gamma$ between the interfering amplitudes.

The decay rate distribution for 
$B \to {\Dstar}^\pm\pi^\mp$ is
\begin{equation}
\label{eq:pure-dt-pdf-B}
\P^\pm_\eta(\dt)
= {e^{-|\dt|/\tau} \over 4\tau} \times       
\left[ 1 \mp S^\zeta \sin(\Delta m \dt)  \mp \eta C \cos(\Delta m \dt) \right], 
\end{equation}
where  
$\tau$ is the $\Bz$ lifetime averaged over the two mass eigenstates,
$\Delta m$ is the $\Bz-\Bzb$ mixing 
frequency, and $\dt$
is the difference between the time
of the $B\to{\Dstar}^\pm\pi^\mp$ ($\Brec$)
decay and the decay of the other
$B$ ($\Btag$) in the event. The
upper (lower) sign in Eq.~(\ref{eq:pure-dt-pdf-B})
indicates the flavor of the $\Btag$ as a $\Bz$ ($\Bzb$),
while $\eta = +1$ ($-1$) and $\zeta = +$ ($-$) for
the $\Brec$ final state ${\Dstar}^-\pi^+$ (${\Dstar}^+\pi^-$).
The parameters $C$ and $S^\pm$ are given by
\begin{equation}
C \equiv {1 - \rsq \over 1 + \rsq}\, , \ \ \ \ 
S^\pm \equiv {2 \r \over 1 + \rsq}\, \sin(2 \beta + \gamma \pm \deltaPhase).
\label{eq:AandB}
\end{equation}
Here $\deltaPhase$ is the strong phase difference 
between $A_u$  and $A_c$ 
and 
$\r = |A_u / A_c|$.
Since $A_u$ is doubly CKM-suppressed with respect
to $A_c$, one expects $\r\sim 0.02$. 

We report a study of the \CP-violating asymmetry in $\btodstpipm$
decays using the technique of partial reconstruction, which allows us
to analyze a large sample of signal events.  
We use approximately twice the integrated luminosity used in our
previous analysis of this process~\cite{ref:run1-2}, 
and employ
an improved method to eliminate a measurement bias,
as described in Sec.~\ref{sec:signalpdf}.

\section{THE \babar\ DETECTOR AND DATASET}
\label{sec:babar}

The data used in this analysis were recorded with the \babar\
detector at the \pep2\ storage ring, and consist of 165.6~fb$^{-1}$
collected on the $\Upsilon(4{\rm S})$ resonance (on-resonance
sample), and 16~fb$^{-1}$ collected at an $\epem$ CM 
energy approximately 40~\mev below the resonance peak 
(off-resonance sample). Samples of simulated Monte Carlo (\mc) events
with an equivalent luminosity $4$  times larger than the data are
analyzed through the same analysis procedure.

The \babar\ detector is described in detail in Ref.~\cite{ref:babar}.
We provide a brief description of the main components and their use in
this analysis.  Charged-particle trajectories are measured by a
combination of a five-layer silicon vertex tracker (SVT) and a
40-layer drift chamber (DCH) in a 1.5 T solenoidal magnetic field.
Tracks with low transverse momentum can be reconstructed in the SVT
alone, thus extending the charged-particle detection down to
transverse momenta of about  50~\mevc. We use a ring-imaging Cherenkov 
detector (DIRC) for charged-particle identification and  augment it with  
energy-loss measurements from the
SVT and DCH. Photons and electrons are
detected in a CsI(Tl) electromagnetic calorimeter (EMC), with 
photon-energy resolution $\sigma_E / E = 0.023 (E/\gev)^{-1/4} \oplus
0.019$. The instrumented flux return (IFR) is equipped with
resistive plate chambers to identify muons.

\section{ANALYSIS METHOD}
\label{sec:Analysis}

\subsection{PARTIAL RECONSTRUCTION OF \boldmath $\btodstpipm$}
\label{sec:partial}

In the partial reconstruction of a $\btodstpipm$ candidate ($\Brec$), 
only the hard (high-momentum) pion track $\pi_h$ from the $B$ decay and the
soft (low-momentum) pion track $\pi_s$ from the decay
$D^{*-}\rightarrow \Dzb \pi_s^-$ are used.
%
%
Applying kinematic constraints consistent with the signal decay mode,
we calculate the four-momentum of the unreconstructed, ``missing''
$D$, obtaining its flight direction to within a few degrees and its
invariant mass $\mmiss$~\cite{ref:dstpi-lifetime}.  Signal events peak
in the $\mmiss$ distribution at the nominal $\Dz$ mass $M_{\Dz}$ with an
r.m.s. of 3~\mevcc (Fig. \ref{fig:data_mmiss}).

\subsection{BACKGROUNDS}
\label{sec:bgd}

In addition to $\btodstpipm$  events, the selected event sample  
contains the following kinds of events:
$\btodstrhopm$; 
$\BB$ background peaking in $\mmiss$,
  composed of pairs of tracks coming from
  the same $B$ meson, with the $\pi_s$ originating from a
  charged $\Dstar$ decay, excluding  $\btodstrhopm$ decays;
combinatoric \BB background, defined as all remaining $\BB$ background 
events;
and continuum $\epem \rightarrow \qqbar$, 
where $q$
represents a $u$, $d$, $s$, or $c$ quark.
We suppress the combinatoric background 
with selection criteria based on the event shape and the 
$\Dstar$ helicity-angle. 
We reject $\pi_h$ candidates
that are identified as leptons or kaons.
All candidates must satisfy the requirement $1.81 < \mmiss <
1.88$~\gevcc. 
Multiple 
candidates are found in 5\% of the events. In these instances,
only the candidate with the $\mmiss$ value closest to $M_{\Dz}$ is used.
%
%

\subsection{DECAY TIME MEASUREMENT AND FLAVOR TAGGING}
\label{sec:deltat}

To perform this analysis, $\dt$ and the flavor of the $\Btag$ must be
determined. 
We tag the flavor of the $\Btag$ using lepton or kaon
candidates.
The lepton CM momentum is required to be greater than 1.1~\gevc 
to suppress ``cascade'' leptons that originate from charm decays.
If several flavor-tagging tracks are present in either the lepton or kaon
tagging category,
the only track of that category used for tagging is the one with the
largest value of $\theta_T$, the CM angle between the track
momentum and the missing $D$ momentum.  The tagging track must satisfy
$\cos \theta_T<C_T $, 
where $C_T=0.75 $ ($C_T=0.50 $) for leptons (kaons),  to minimize
the impact of tracks originating from the $D$ decay. If both a lepton and a
kaon satisfy this requirement, the event is tagged with the lepton only.

We measure $\dt$ using $\dt = (\zrec - \ztag) /
(\gamma\beta c)$, where $\zrec$ ($\ztag$) is the decay position of the
$\Brec$ ($\Btag$) along the beam axis ($z$) in the laboratory frame,
and the $e^+e^-$ boost parameter $\gamma\beta$ is
continuously calculated from the beam energies. 
To find $\zrec$, we use the $\pi_h$ track parameters and errors,
and the beam-spot position and size in the plane perpendicular to the
beams (the $x - y$ plane). We find the position of the point in space
for which the sum of the $\chi^2$ contributions from the $\pi_h$ track
and the beam spot is a minimum. The $z$ coordinate of this point
determines $\zrec$.
In lepton-tagged events, the same procedure,
with the $\pi_h$ track replaced by the tagging lepton, is used to
determine $\ztag$.

In kaon-tagged events, we obtain $\ztag$ from a beam-spot-constrained
vertex fit of all tracks in the event, excluding $\pi_h$, $\pi_s$ and all tracks
within 1~rad of the $D$ momentum in the CM frame.
If the contribution of any track to the $\chi^2$ of the vertex 
is more than 6, the track is removed and the fit is repeated until
no track fails the $\chi^2 < 6$ requirement.

The $\dt$ error $\dtErr$ is calculated from the results
of the $\zrec$ and $\ztag$ vertex fits.

\subsection{PROBABILITY DENSITY FUNCTION}
\label{sec:pdf}

The analysis is carried out with a series of unbinned
maximum-likelihood fits, performed simultaneously on the on- and
off-resonance data samples and independently for the lepton-tagged and
kaon-tagged events.
The probability density function (PDF) depends on the variables
$\mmiss$, $\dt$, $\dtErr$, $F$, $\stag$, and $\smix$,
where 
$F$ is a Fisher discriminant formed from fifteen event-shape variables
that provide discrimination against continuum
events~\cite{ref:dstpi-lifetime},
$\stag = 1$ ($-1 $) when the $\Btag$ is identified as a $\Bz$ ($\Bzb$), 
and $\smix = 1$ ($-1 $) for ``unmixed'' (``mixed'') events.
An event is labeled unmixed if the $\pi_h$ is a 
$\pi^- (\pi^+)$ and the $\Btag$ is a $\Bz (\Bzb)$, and mixed
otherwise.

The PDF for on-resonance data is a sum over the PDFs of
the different event types:
\begin{equation}
\P = \sum_{i}f_{i} \, \P_i,
\label{eq:pdf-sum}
\end{equation}
where the index $i = \{\dstpi, \dstrho, \peak, \comb, \cont\}$
indicates one of the event types described above, $f_i$ is the
relative fraction of events of type $i$ in the data sample, and $\P_i$
is the PDF for these events.
The PDF for off-resonance data is $\P_\cont$.
The parameter values for $\P_i$ are different for each event type,
unless indicated otherwise.  Each $\P_i$ is a product,
\beq 
\P_i = \M_i(\mmiss)\, \F_i(F)\, \T'_i(\dt, \dtErr, \stag, \smix),
\label{eq:pdf-prod}
\eeq
where the factors in Eq.~(\ref{eq:pdf-prod}) are described below.

\subsubsection{\boldmath $\mmiss$ AND $F$ PDFs }

The $\mmiss$ PDF for each event type $i$ is the sum of a bifurcated
Gaussian plus an ARGUS function:
\begin{equation}
\M_i(\mmiss) = f^{\BG}_i\, \BG_i(\mmiss) + (1-f^{\BG}_i) \A_i(\mmiss), 
\label{eq:mmiss-pdf}
\end{equation}
where $f^{\BG}_i$ is the bifurcated Gaussian fraction. The functions 
$\BG_i$ and $\A_i$ are
\begin{eqnarray}
\BG_i(x) &\propto& \biggl\{ \matrix{
        \exp\left[-(x - M_i)^2 / 2\sigma_{Li}^2\right] & , & 
                                        x < M_i \cr
        \exp\left[-(x - M_i)^2 / 2\sigma_{Ri}^2\right] & , & 
                                        x > M_i 
                            }, 
\label{eq:bifur}
\\[0.2cm]
 \A(x) 
        &\propto& \biggl\{ \matrix{x \sqrt{1-\left({x /M^A_i}\right)^2}\;
   \exp\left[\epsilon_i \left(1-\left({x / M^A_i}\right)^2\right)\right] 
         & , & x < M^A_i \cr
	0& , & x \ge M^A_i,
	},
\label{eq:argus}
\end{eqnarray}
where $M_i$ is the peak of the bifurcated Gaussian, $\sigma_{Li}$ and
$\sigma_{Ri}$ are its left and right widths, $\epsilon_i$ is
the ARGUS exponent, $M^A_i$ is its end point, and the proportionality
constants are such that each of these functions is normalized to unit
area. 
The $\mmiss$ PDF of each event type has different parameter values.


The Fisher discriminant PDF $\F_i$ for each event type is
parameterized as a bifurcated Gaussian, as in Eq.~(\ref{eq:bifur}). The
parameter values of $\F_\dstpi$, $\F_\dstrho$, $\F_\peak$, and
$\F_\comb$ are identical.

\subsubsection{\boldmath SIGNAL  $\dt$ PDFs}
\label{sec:signalpdf}

The $\dt$ PDF $\T'_\dstpi(\dt, \dtErr, \stag, \smix)$ for signal events
corresponds to Eq.~(\ref{eq:pure-dt-pdf-B}) with $O(\rsq)$ 
terms neglected, and with additional parameters that account
for several experimental effects, described below.

The first effect has to do with the origin of the tagging track.
In some of the events, the tagging track
originates from the decay of the missing $D$.
These events are labeled ``missing-$D$ tags'' and do not 
provide any information regarding the flavor of the $\Btag$.
In lepton tag events we further distinguish between ``direct'' tags,
in which the tagging lepton originates directly from the decay of the
$\Btag$, and ``cascade'' tags, where the tagging lepton is a daughter
of a charmed particle produced in the $\Btag$ decay. Cascade and
direct tags have different mistag probabilities due to the different
physical origin of the tagging track.  In addition,  the measured value
of $\ztag$ in cascade-lepton tags is systematically larger than the
true value, due to the finite lifetime of the charmed particle and the
boosted CM frame.  This creates a correlation between the tag and
vertex measurements that we address by considering
cascade-lepton tags separately in the PDF~\footnote{ 
  In Ref.~\cite{ref:run1-2} we corrected for the bias caused by this
  effect and included a systematic error due to its uncertainty.}.
In kaon tags, $\ztag$ is determined using all available $\Btag$
tracks.  Therefore, the effect of the tagging track on the $\ztag$
measurement is small, and no distinction between cascade and direct
kaon tags is needed.

The second experimental effect is the finite detector resolution in the
measurement of $\dt$. We address this by convoluting the distribution
of the true decay time difference $\dttrue$ with a detector resolution
function. Putting these two effects together, the $\dt$ PDF of signal
events is
\beq
\T'_\dstpi (\dt, \dtErr, \stag, \smix) = (1 + \Delta\epsilon_\dstpi \, \stag) 
	\sum_j f^j_\dstpi
        \int d\dttrue\, \T_\dstpi^j(\dttrue, \stag, \smix) \,
        \R_\dstpi^j(\dt - \dttrue, \dtErr),
\label{eq:CP-pdf-sig}
\eeq
where
$2 \Delta\epsilon_\dstpi$ is the relative difference between 
the detection efficiencies of positive and negative leptons or kaons,
the index $j = \{\dir, \ \cas, \ \mis\}$ indicates direct, cascade, and
missing-$D$ tags,
and $f^j_\dstpi$ is the fraction of signal events of tag-type $j$ in
the sample.  We set $\fdir_\dstpi = 1 - \fcas_\dstpi -
\fmis_\dstpi$. For kaon tags $\fcas_\dstpi = 0$.
The function $\T_\dstpi^j(\dttrue, \stag, \smix)$ is the $\dttrue$ distribution
of tag-type $j$ events,
and $\R_\dstpi^j(\dt - \dttrue, \dtErr)$ 
is their resolution function, which parameterizes 
both the finite detector resolution and systematic 
offsets in the measurement of $\dz$, such as those due to the 
origin of the tagging particle. The parameterization of the resolution
function is described in Sec.~\ref{sec:res}. 


The $\dttrue$ PDF for missing-$D$ tags is
\beq
\Tmis_\dstpi(\dttrue, \smix)= 
	{e^{-|\dttrue|/\tau^\mis_\dstpi}\over 8 \tau^\mis_\dstpi}  
	\left\{1 + \smix \left( 1 - 2\rho_\dstpi\right) \right\},
\label{eq:CP-pdf-mis}
\eeq
where $\rho_\dstpi$ is the probability that the charge of the tagging track
is such that it results in a mixed flavor measurement.

The functional form of the direct and cascade tag $\dttrue$ PDFs is
\bea
\T^j_\dstpi(\dttrue, \stag, \smix)
	= {e^{-|\dttrue|/\tau_\dstpi}\over 4 \tau_\dstpi}        
        \biggl\{&&  \kern-0.6cm
		1-\stag\,\Delta\omega^j_\dstpi
	\nonumber\\
        &&  \kern-0.6cm + \smix\,(1-2\omega^j_\dstpi)\, \cos(\Delta m_\dstpi
          \dttrue)      
	\nonumber\\   
        &&  \kern-0.6cm  - {\cal S}^j_\dstpi 
                \,\sin(\Delta m_\dstpi \dttrue) 
         \biggr\},
\label{eq:CP-pdf-dir-cas}
\eea
where 
$j = \{\dir, \ \cas\}$, the mistag rate $\omega^j_\dstpi$ is the probability to
misidentify the flavor of the $\Btag$ averaged over $\Bz$ and $\Bzb$,
and $\Delta \omega^j_\dstpi$ is the 
$\Bz$ mistag rate minus the $\Bzb$ mistag rate.
The factor ${\cal S}^j_\dstpi$ describes CP violation due to interference 
between $\btou$ and $\btoc$ amplitudes in both 
the $\Brec$ and the 
$\Btag$ decays:
\beq
{\cal S}^j_\dstpi=  (1-2\omega^j_\dstpi) \, (\stag a_\dstpi + \smix c_\dstpi) 
        + \stag \smix b_\dstpi (1-\stag \Delta\omega^j_\dstpi) ,
\label{eq:cp-pdf-sin}
\eeq
where $a_\dstpi$, $b_\dstpi$, and $c_\dstpi$ are related to the physical
parameters through
\bea
a_\dstpi&\equiv& 2 \r\sin(2\beta+\gamma)\cos\deltaPhase , \nonumber\\
b_\dstpi&\equiv& 2 \rp\sin(2\beta+\gamma)\cos\deltaPhaseP , \nonumber\\
c_\dstpi&\equiv& 2\cos(2\beta+\gamma)(\r\sin\deltaPhase -\rp\sin\deltaPhaseP), 
\label{eq:abc}
\eea
and $\rp$ ($\deltaPhaseP$) is the effective magnitude of the ratio 
of amplitudes (strong phase
difference) between the $b\rightarrow u \overline c
d$ and $b\rightarrow c \overline u d$ amplitudes in the $\Btag$ 
decay. 
This parameterization is good to first order in $\r$ and $\rp$.
The inclusion of $\rp$ and $\deltaPhaseP$ in the 
formalism accounts for cases where the 
$\Btag$ undergoes a $b\rightarrow u \bar c d$ decay, and the kaon 
produced in the subsequent charm decay is used for tagging~\cite{ref:abc}.
In lepton-tagged events $\rp = 0$ (and hence $b_\dstpi=0$).

\subsubsection{\boldmath BACKGROUND  $\dt$ PDFs}

The $\dt$ PDF of $\btodstrhopm$ has the same functional form and
parameter values as the signal PDF, except that the \CP parameters
$a_\dstrho$, $b_\dstrho$, and $c_\dstrho$ are set to 0 and are later
varied to evaluate systematic uncertainties.
The validity of the use of the same non-CP parameters for 
$\T'_\dstrho$ and 
$\T'_\dstpi$ is established using simulated events, and
stems from the fact that the $\pi_h$ momentum spectrum in the 
$\btodstrhopm$  events that pass our selection criteria is almost 
identical
to the signal spectrum.

The $\dt$ PDF of the peaking background accounts separately for 
charged and neutral $B$ decays:
\bea
\T'_\peak (\dt, \dtErr, \stag, \smix) &=& (1 + \stag \, \Delta\epsilon_\peak)
	\nonumber\\
	&\times&
	\left\{\T{^0}'_\peak +
        \int d\dttrue\, \T_\peak^+(\dttrue, \stag, \smix) \,
        \R_\peak^+(\dt - \dttrue, \dtErr) \right\},	
\label{eq:pdf-peak}
\eea
where $\T{^0}'_\peak$ has the functional form of
Eq.~(\ref{eq:CP-pdf-sig}) and the subsequent expressions,
Eqs.~(\ref{eq:CP-pdf-mis}-\ref{eq:abc}), but with all $\dstpi$-subscripted
parameters replaced with their $\peak$-subscripted counterparts. 
The integral in Eq.~(\ref{eq:pdf-peak}) accounts for 
the contribution of charged $B$ decays to the peaking background,
with 
\beq
\T_\peak^+(\dttrue, \stag) = 
	{e^{-|\dttrue|/\tau^+_\peak}\over 4 \tau^+_\peak}        
        \left(1-\stag\,\Delta\omega^+_\peak \right),
\label{eq:pdf-peak-charged}
\eeq
and $\R_\peak^+(\dt - \dttrue, \dtErr)$ being the 3-Gaussian resolution 
function for these events.
%

Convergence of the fit with the parameterization of $\T'_\dstpi$
relies on our ability to independently determine $\fmis_\dstpi$ 
using the angle between the $\pi_s$ and the $\Dstar$,   
as described later in this note. 
That determination may not be carried out reliably 
for the combinatoric $\BB$ background, due uncertainties in 
simulating the random
nature of the reconstructed track combinations.
As a result, the four parameters
$\fdir_\comb$, $\omega^\dir_\comb$, $\Delta\omega^\dir_\comb$, 
and $\rho_\comb$ are not independent, and can be reduced to the
set of three parameters
\begin{eqnarray}
  \omega'_\comb &= & 
     \omega_\comb^\dir\,(1-\fdir_\comb)+\frac{\fdir_\comb}{2}, \nonumber\\
  \Delta\omega'_\comb &= &  
      \Delta\omega_\comb\,(1-\fdir_\comb),  \nonumber\\
  {\Omega_\comb} & = & \fdir_\comb (1-2\,\rho_\comb). 
\end{eqnarray}
With these parameters and $\fcas_\comb = 0$, the 
combinatoric $\BB$ background $\dt$ PDF becomes 
\beq
\T'_\comb (\dt, \dtErr, \stag, \smix) =  (1 + \stag \, \Delta\epsilon_\comb)
        \int d\dttrue\, \T_\comb(\dttrue, \stag, \smix) \,
        \R_\comb(\dt - \dttrue, \dtErr) ,	
\label{eq:pdf-comb}
\eeq
where $\R_\comb(\dt - \dttrue, \dtErr)$ is the 3-Gaussian resolution fucntion
and
\bea
\T_\comb(\dttrue, \stag, \smix)= {e^{-|\dttrue|/\tau_\comb}\over 4 \tau_\comb} 
	\biggl\{&&  \kern-0.6cm 1-\stag\,\Delta\omega'_\comb 
	+ \smix \Omega_\comb
	  \nonumber\\        
	 &&\kern-0.6cm  
         + \smix\,(1-2\omega'_\comb)\, \cos(\Delta m_\comb
          \dttrue)         
	\nonumber\\
         &&\kern-0.6cm 
	- {\cal S}_\comb 
                \,\sin(\Delta m_\comb \dttrue) 
         \biggr\},
\label{eq:CP-pdf-comb}
\eea
with
\beq
{\cal S}_\comb=  (1-2\omega'_\comb) \, (\stag a_\comb + \smix c_\comb) 
        + \stag \smix b_\comb (1-\stag \Delta\omega'_\comb) .
\label{eq:cp-pdf-sin-comb}
\eeq

As in the case of $T_\dstrho$, the $CP$ parameters $a_\peak$, $b_\peak$,
and $c_\peak$ are set to 0
and are later varied to evaluate systematic uncertainties.
Parameters labeled with superscripts ``$\peak$'' or ``$\comb$''
are empirical and thus do not necessarily correspond to 
physical parameters and may have different values from those of 
the $\dstpi$-labeled parameters.

%
The PDF $\T_\cont$ for the continuum background is
the sum of two components, one with a finite lifetime
and one with zero lifetime:
\beq
\T'_\cont (\dt, \dtErr, \stag) = (1+\stag \, \Delta\epsilon_\cont)
        \int d\dttrue\, \T_\cont(\dttrue, \stag, \smix) \,
        \R_\cont(\dt - \dttrue, \dtErr) ,	
\label{eq:pdf-qq}
\eeq
with 
\beq
\T_\cont(\dttrue, \stag) = (1-f^\delta_\cont)
	{e^{-|\dttrue|/\tau_\cont}\over 4 \tau_\cont}        
        \left(1-\stag\,\Delta\omega_\cont \right)
	 + f^\delta_\cont\, \delta(\dttrue),
\label{eq:pdf-qq-2}
\eeq
where $f^\delta_\cont$ is the fraction of zero-lifetime events.

\subsubsection{RESOLUTION FUNCTION PARAMETERIZATION}
\label{sec:res}

The resolution function for events of type $i$ and optional secondary-type
$j$ ($j = \{\dir, \ \cas, \ \mis\}$ for lepton-tagged signal events and $j = \{+, \ 0\}$ 
for the $\BB$ background types)
is parameterized as the sum of
three Gaussians:
\begin{equation}
\R\ij(\dt - \dttrue, \dtErr) = 
        f^n\ij\, \G^n\ij(t_r, \dtErr) 
	+ (1 - f^n\ij - f^o\ij)\, \G^w\ij(t_r, \dtErr) 
        + f^o\ij\, \G^o\ij(t_r, \dtErr),
\label{eq:res}
\end{equation}
where $t_r$ is the residual $\dt - \dttrue$, and $\G^n\ij$, 
$\G^w\ij$, and
$\G^o\ij$ are the ``narrow'', ``wide'', and ``outlier'' Gaussians. The
narrow and wide Gaussians have the form
\begin{equation}
\G^k\ij(t_r, \dtErr) \equiv 
        {1 \over \sqrt{2\pi} \, s^k\ij\, \dtErr}
  \exp\left(-\,{\left(t_r - b^k\ij\dtErr\right)^2  
        \over 2 (s^k\ij\, \dtErr)^2}\right), 
\label{eq:Gaussians}
\end{equation}
where the index $k$ takes the values $k=n,w$ for the narrow and wide
Gaussians, and $b^k\ij$ and $s^k\ij$ are parameters determined by
fits, as described in Sec.~\ref{sec:proc}.
The outlier Gaussian has the form 
\begin{equation}
\G^o\ij(t_r, \dtErr) \equiv 
        {1 \over \sqrt{2\pi} \, s^o\ij  }
  \exp\left(-\,{\left(t_r - b^o\ij \right)^2  
        \over 2 (s^o\ij)^2}\right),
\label{eq:GaussiansOutlier}
\end{equation}
where in all nominal fits the values of $b^o\ij$ and $s^o\ij$ are
fixed to 0~ps and 8~\ps, respectively, and are later varied to
evaluate systematic errors.

\subsection{ANALYSIS PROCEDURE}
\label{sec:proc}

The analysis proceeds in four steps involving unbinned maximum likelihood
fits to the data:
\begin{enumerate}
\item In the first step, we determine the parameters
$f_\dstrho+f_\dstpi$, $ f_\peak$, and 
$f_\comb$ of Eq.~(\ref{eq:pdf-sum}).
In order to reduce the reliance on the simulation, we also obtain in
the same fit the parameters
$f^{\BG}_\cont$ of Eq.~(\ref{eq:mmiss-pdf}), 
$\epsilon_\cont$ of Eq.~(\ref{eq:argus}), $\sigma_{L}$ for the signal (Eq. \ref{eq:bifur})
and all the parameters of $\F_\cont$, and  $\F_{\BB}$ (the latter applies
to all $\BB$ event types). This is done by fitting the data with 
the PDF
\beq
\P_i = \M_i(\mmiss) \, \F_i(F),
\label{eq:pdf-prod-kin}
\eeq
instead of Eq.~(\ref{eq:pdf-prod}), i.e. by ignoring the time dependence.  
The fraction $f_\cont$ of continuum events is determined
from the event yield of the off-resonance sample and its 
integrated luminosity relative to the on-resonance sample.
All other parameters of the $\M_i$ PDFs and the value 
of $f_\dstpi/(f_\dstpi + f_\dstrho)$ are obtained
from the \mc\ simulation.

\item In the second step, we repeat the fit of the first step for data
events with $\cos \theta_T \ge C_T$, to obtain the fraction of signal
events in that sample.  Given this fraction and the relative
efficiencies for direct, cascade, and missing-$D$ signal events to
satisfy the $\cos \theta_T<C_T$ requirement, we calculate
$\fmis_\dstpi$.  We also calculate
the value of $\rho_\dstpi$ from the fractions of mixed and unmixed signal
events in the $\cos \theta_T \ge C_T$ sample relative to the $\cos
\theta_T < C_T$ sample.

\item In the third step, we fit the data events in the sideband $1.81 <
\mmiss < 1.84$~\gevcc with the 3-dimensional PDFs of
Eq.~(\ref{eq:pdf-prod}).  The parameters of $\M_i(\mmiss)$ and $\F_i(F)$,
and the fractions $f_i$ are fixed to the values obtained in the
first step.  
From this fit we obtain the parameters of $\T'_{\comb}$, as well as
those of $\T'_{\cont}$.

\item In the fourth step, we fix all the parameter
values obtained in
the previous steps and fit the events in the signal region
$1.845 < \mmiss < 1.880$~\gevcc, determining the parameters
of $\T'_\dstpi$ and $\T'_{\cont}$.
Simulation studies show that the parameters of $\T'_{\comb}$ are
independent of $\mmiss$, enabling us to obtain them in the sidebad fit
(step~3) and then use them in the signal-region fit.
The same is not true of the $\T'_{\cont}$ parameters; hence they
are free parameters in the signal-region fit of the last step.
\end{enumerate}


\section{RESULTS}

\begin{figure}[!htbp]
\begin{center}
\begin{tabular}{cc}
   \includegraphics[width=0.48\textwidth]{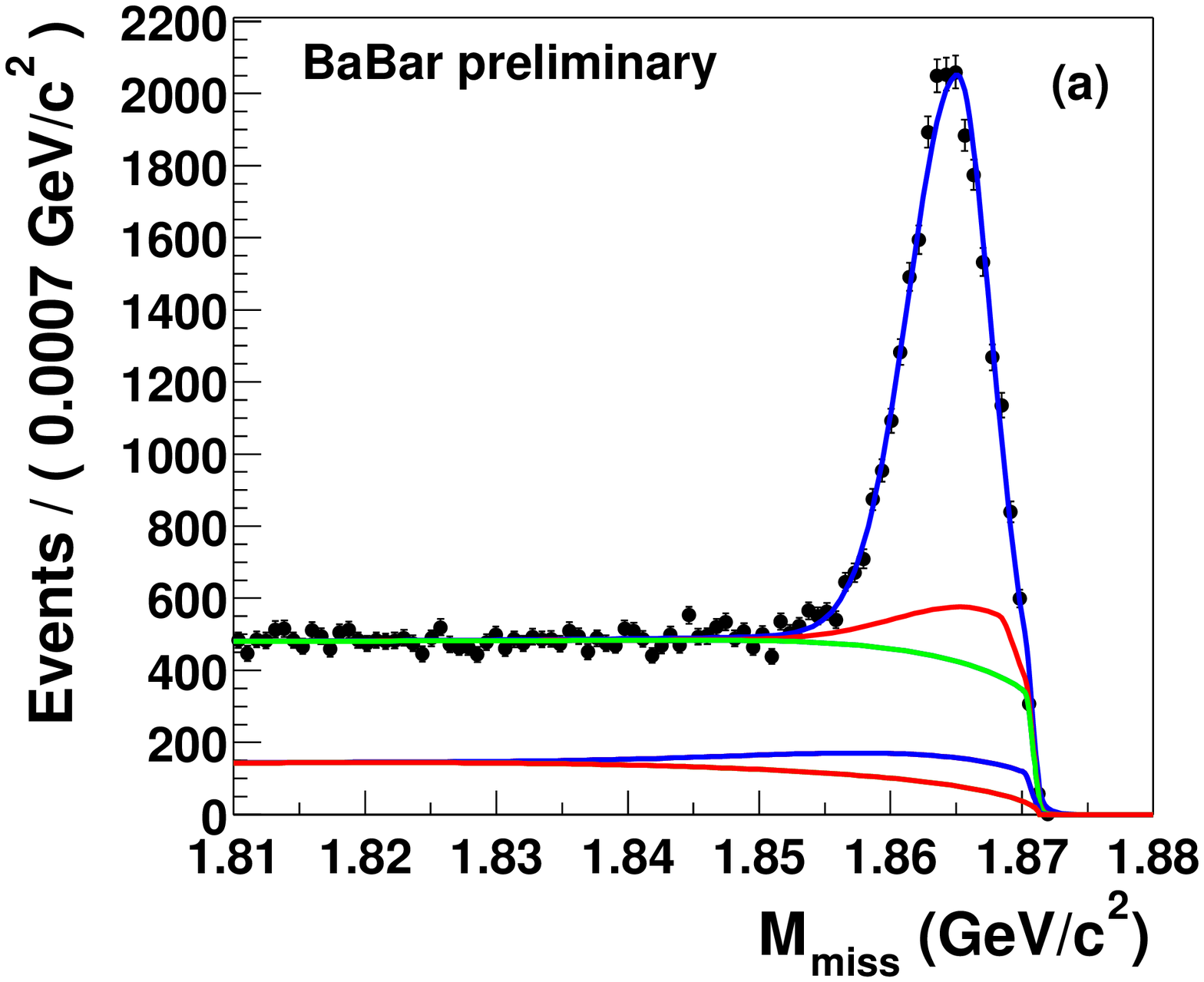}
&
   \includegraphics[width=0.48\textwidth]{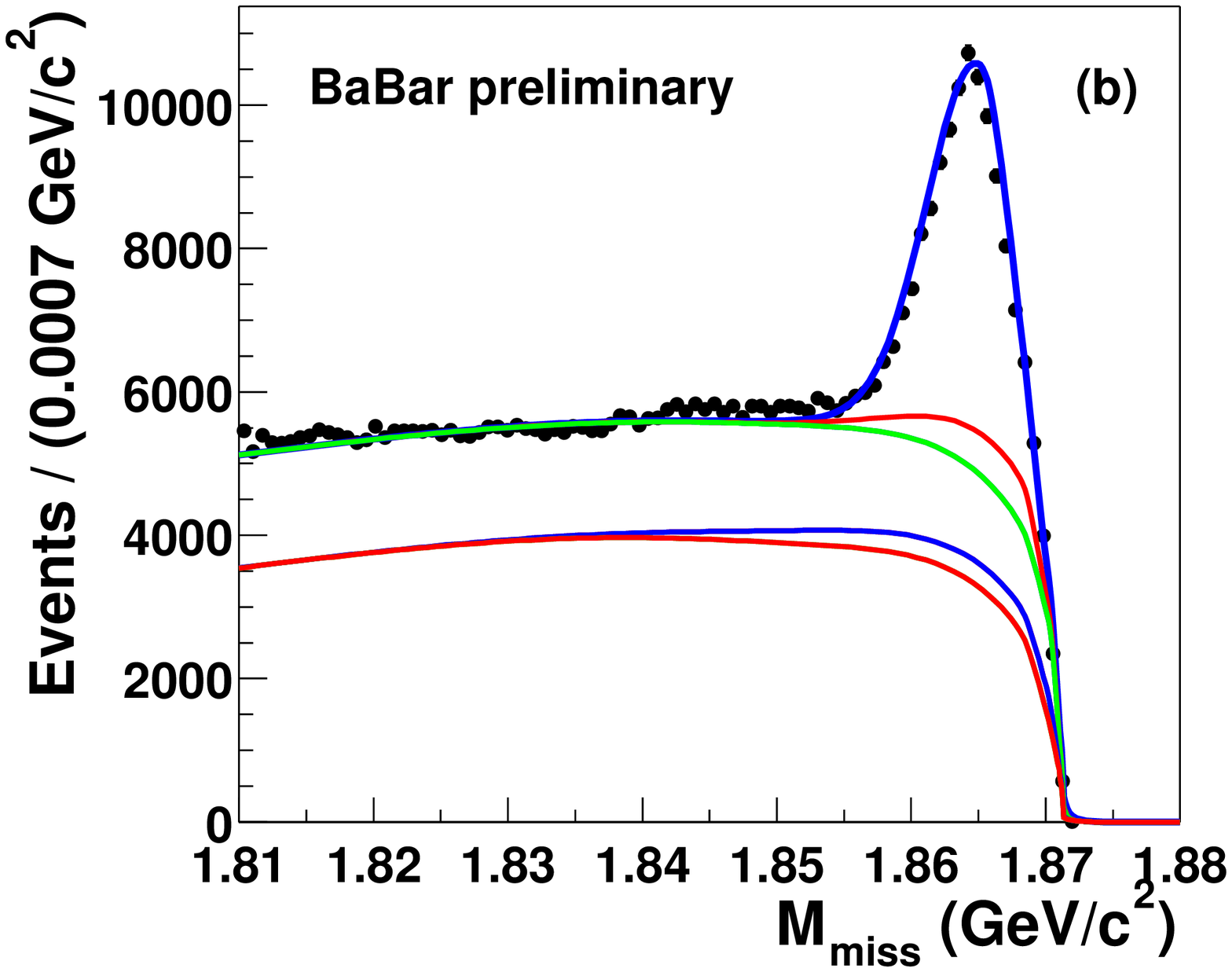}
\end{tabular}
\end{center}
\caption{The $\mmiss$ 
distributions for on-resonance lepton-tagged
(left) and kaon-tagged (right) data. 
The curves show, from bottom
to top, the cumulative contributions of the continuum, peaking \BB, 
combinatoric \BB, $\btodstrhopm$,
and $\btodstpipm$ PDF components.
}
\label{fig:data_mmiss}
\end{figure}
\begin{figure}[!htbp]
\begin{center}
\begin{tabular}{cc}
   \includegraphics[width=0.48\textwidth]{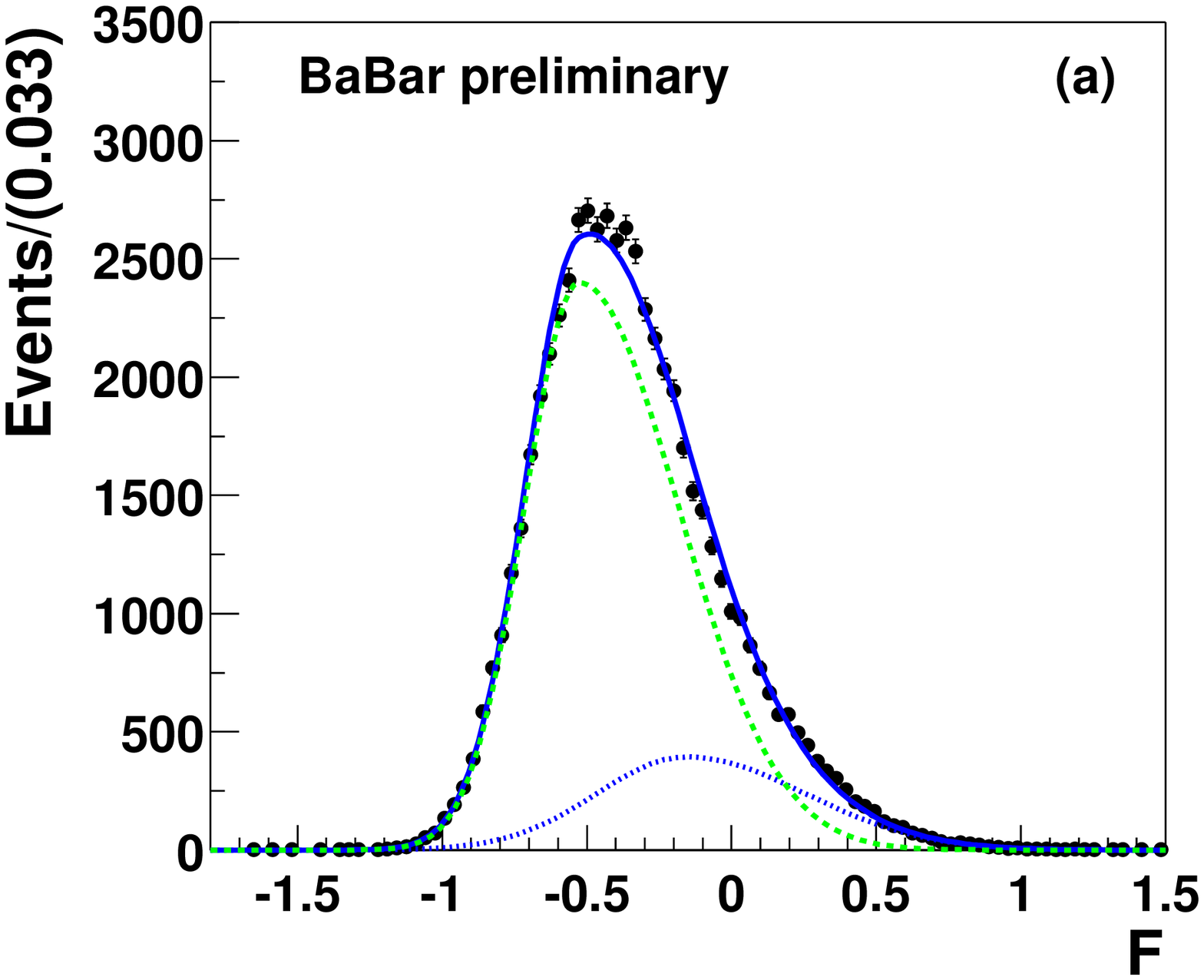}
&
   \includegraphics[width=0.48\textwidth]{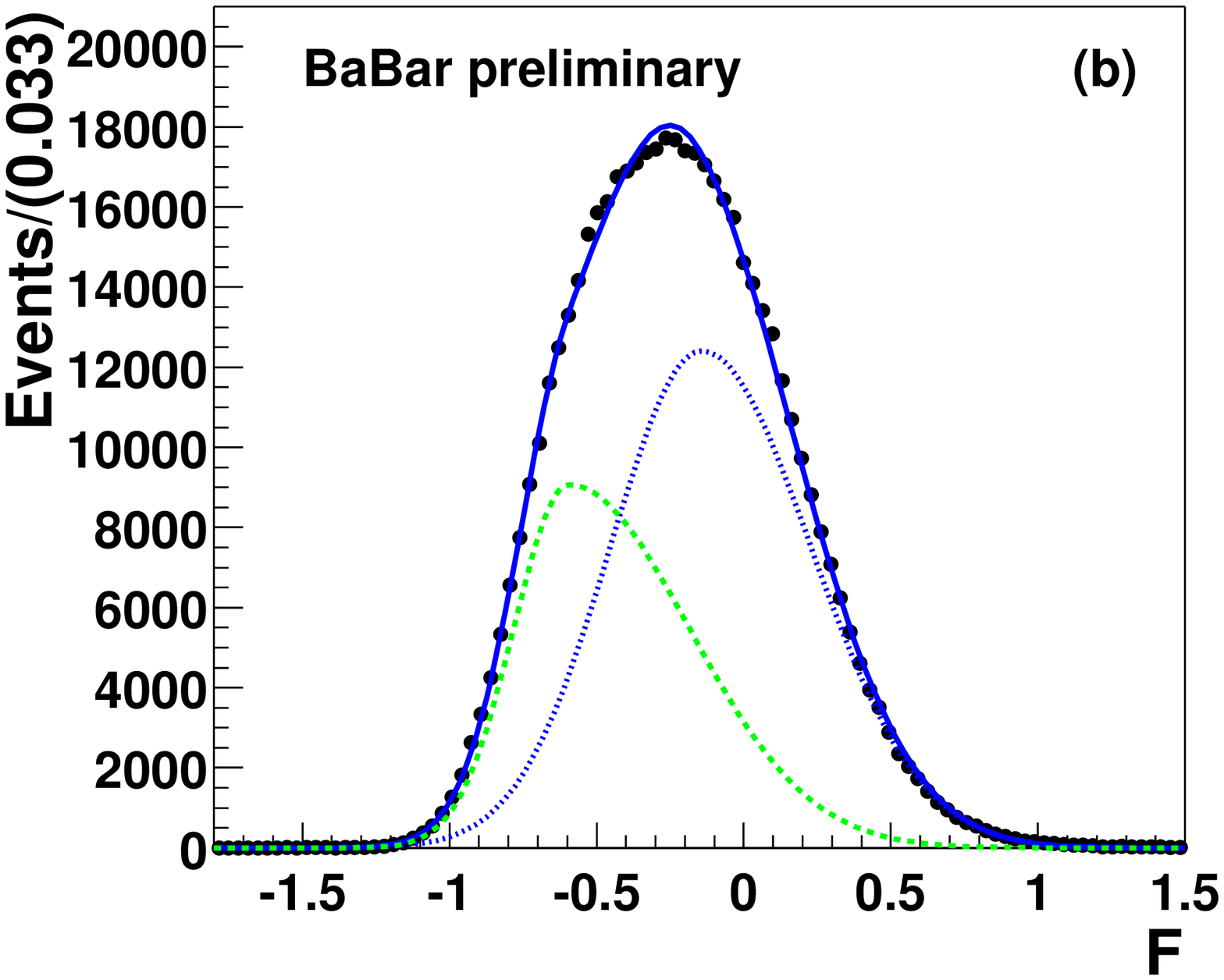}
\end{tabular}
\end{center}
\caption{The $F$ 
distributions for on-resonance lepton-tagged
(left) and kaon-tagged (right) data. 
The contributions of the $\BB$ (dashed line) and the continuum (dot
line) PDF components
are overlayed,
peaking at approximately $-0.6$ and $-0.1$, respectively.
Also overlayed is the total PDF.
}
\label{fig:data_fisher}
\end{figure}

The fit of step~1 finds $16060 \pm 210$ signal $\btodstpipm$ events in
the lepton-tag category and $57480 \pm 540$ in the kaon-tag category.  The
$\mmiss$ and $F$ distributions for data are shown in
Figs.~\ref{fig:data_mmiss} and~\ref{fig:data_fisher}, with the PDFs
overlayed.

\begin{figure}[!htb]
\begin{center}
\includegraphics[width=\textwidth]{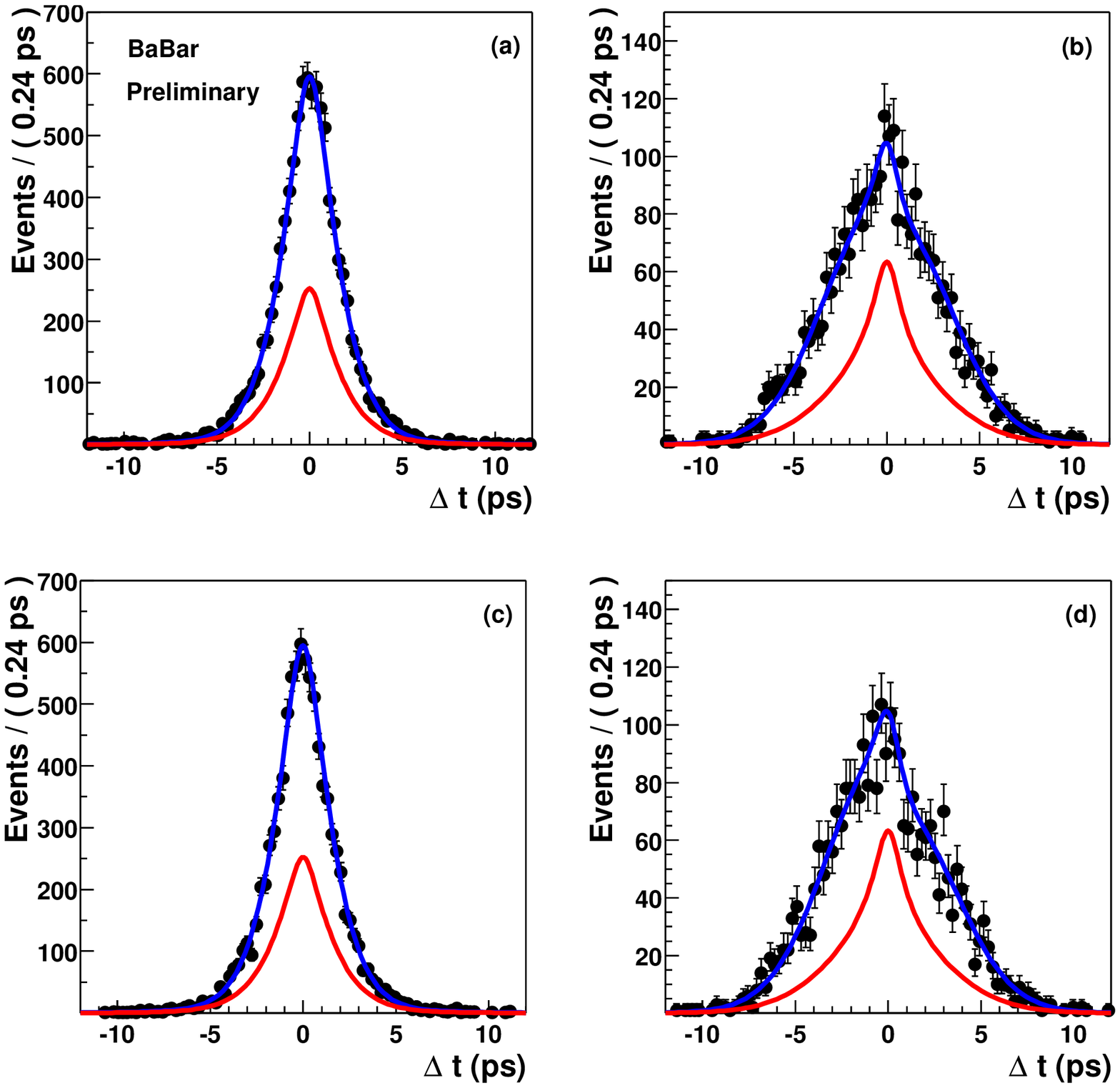}
\end{center}
\caption{$\dt$ distributions for the lepton-tagged events 
separated according to the tagged flavor of $\Btag$ and 
whether they were found to be mixed or unmixed:
a) $\Bz$ unmixed 
%
b) $\Bz$ mixed 
%
c) $\Bzb$ unmixed, 
%
d) $\Bzb$ mixed. 
The curves show the PDF, calculated with the parameters obtained by the fit.
Also shown is the PDF of the total background.
The requirements $\mmiss> 1.855$~\gevcc, $F < 0$ are applied in 
order
to reduce the background.
}
\label{fig:datasr_leptag-Run1-4_log}
\end{figure}

\begin{figure}
\begin{center}
        \includegraphics[width=0.9\textwidth]{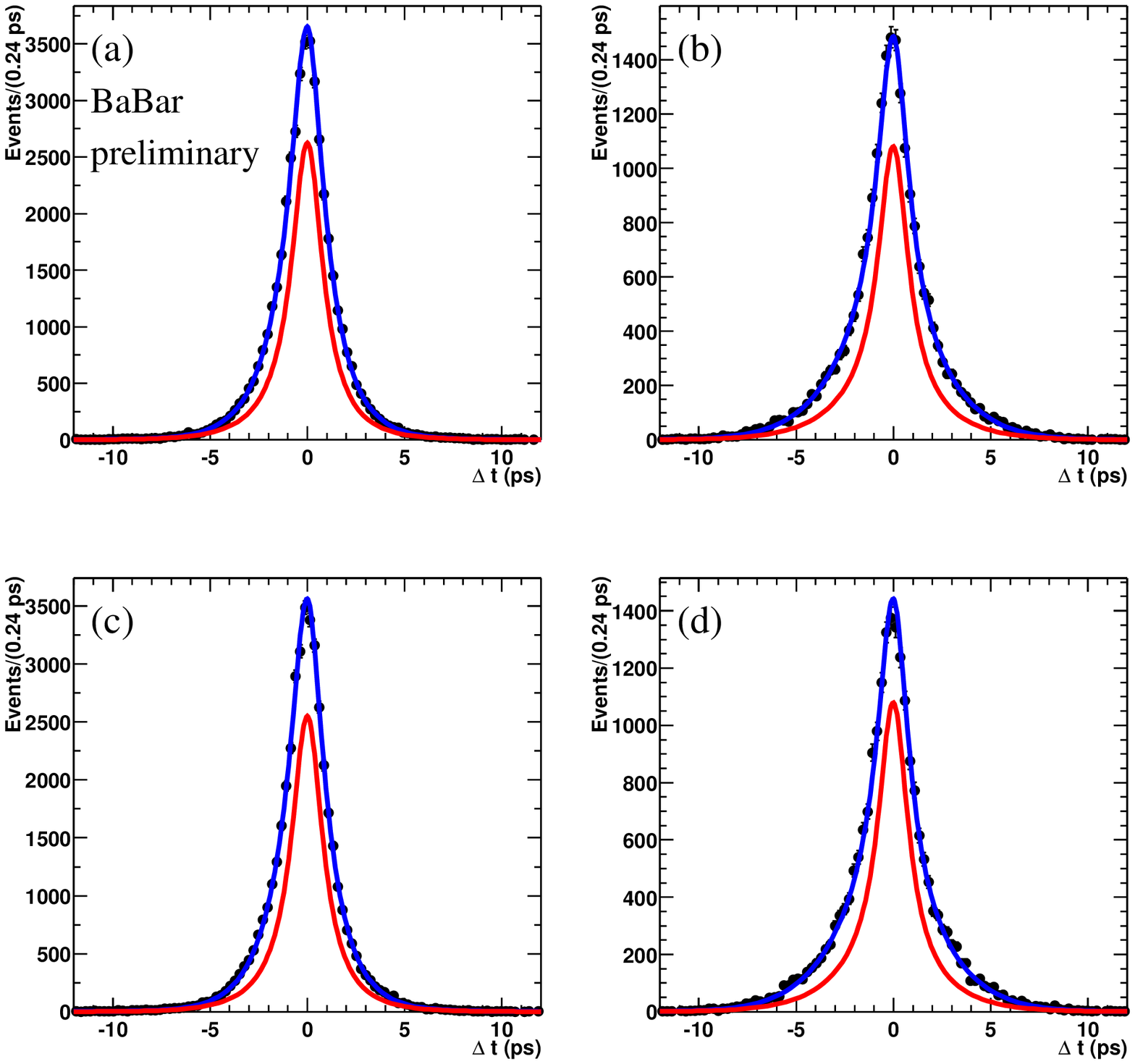}
\end{center}
\caption{$\dt$ distributions for the kaon-tagged events 
separated according to the tagged flavor of $\Btag$ and 
whether they were found to be mixed or unmixed:
a) $\Bz$ unmixed 
%
b) $\Bz$ mixed 
%
c) $\Bzb$ unmixed, 
%
d) $\Bzb$ mixed. 
The curves show the PDF, calculated with the parameters obtained by the fit.
Also shown is the PDF of the total background.
The requirements $\mmiss> 1.855$~\gevcc, $F < 0$ are applied in order
to reduce the background.
}
\label{fig:r1-4-kaon-log}
\end{figure}

\begin{figure}
\begin{center}
\begin{tabular}{cc}
   \includegraphics[width=0.48\textwidth]{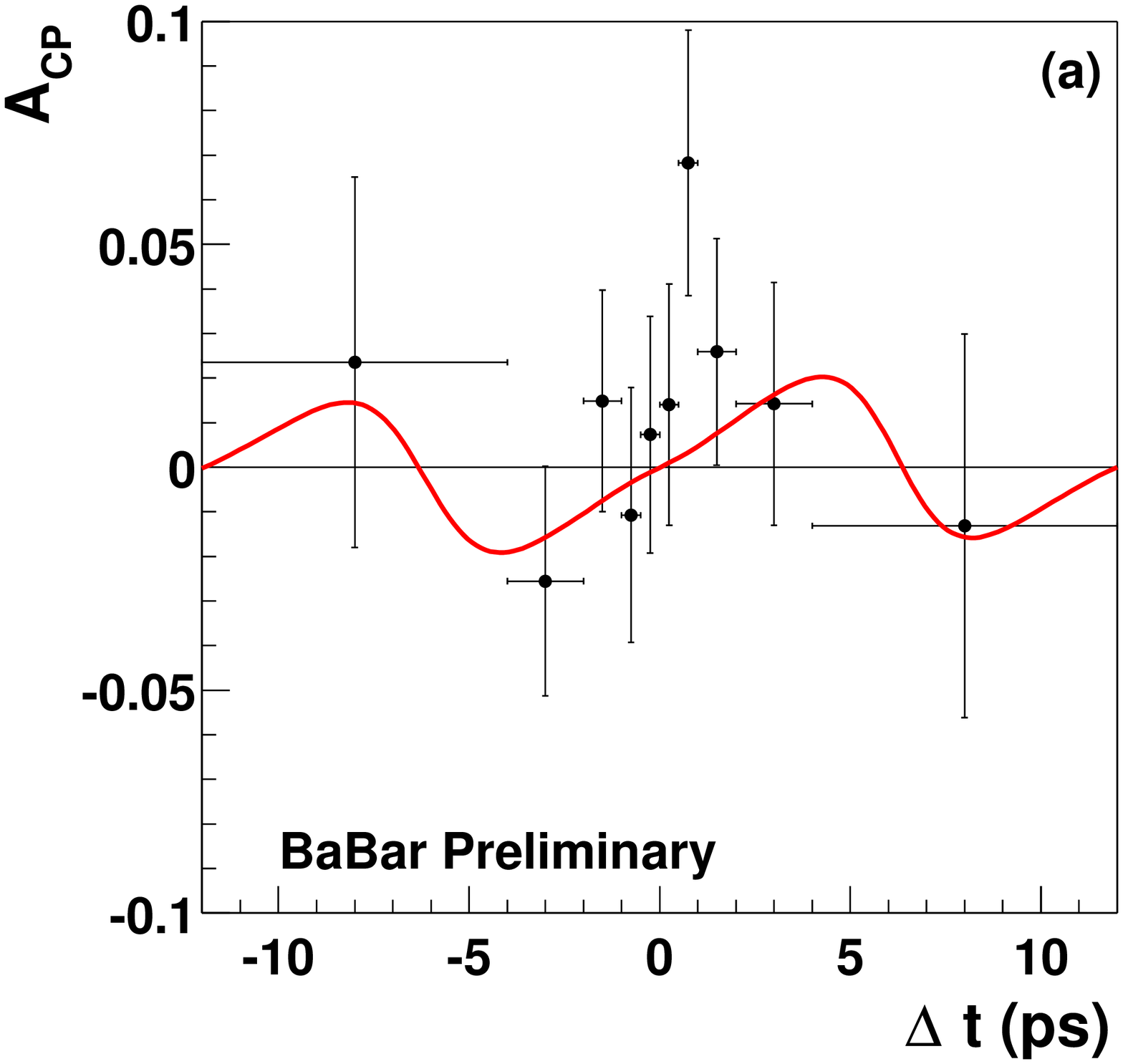}
&
   \includegraphics[width=0.48\textwidth]{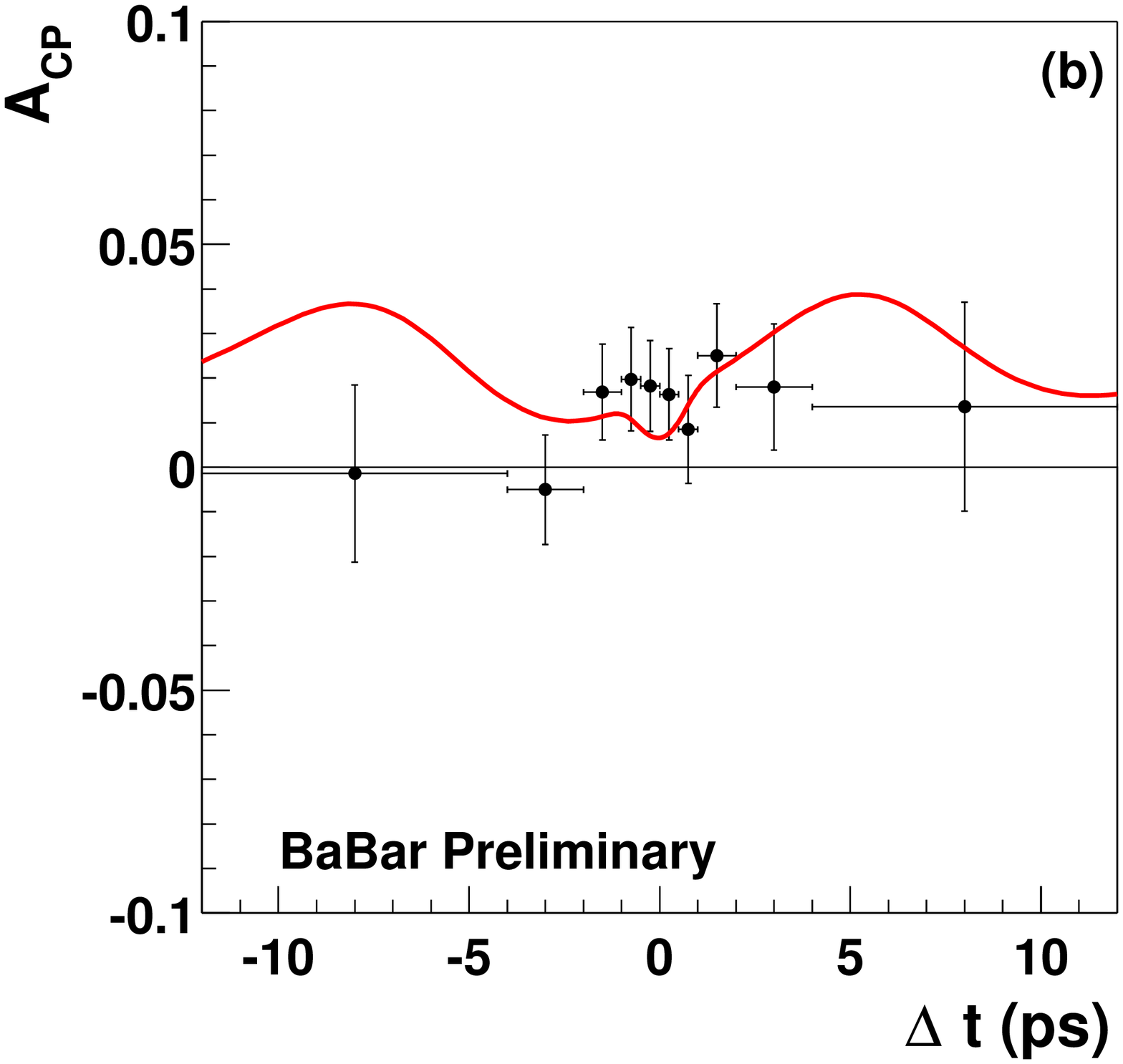}
\end{tabular}
\end{center}
\caption{Raw asymmetry for (a) lepton-tagged and (b) kaon-tagged
  events. The curve represents  the projection of the PDF for the raw
  asymmetry. 
The requirements $\mmiss> 1.855\gevcc$, $F < 0$ are applied in order
to reduce the background. A non-zero value of $a_\dstpi$
would show up as a sinusoidal asymmetry, up to resolution 
and background effects. The offset from the horizontal axis
in the kaon-tag plot is due to the nonzero value of $\Delta \epsilon_{\dstpi}$.}
\label{fig:asym}
\end{figure}

The results of the signal region fit for lepton-tagged events are summarized in
Table~\ref{tab:data_sr_leptag-Run1-4}, and the plots of the $\dt$
distributions for the data are shown in
Fig.~\ref{fig:datasr_leptag-Run1-4_log}.
Results of the fit for the kaon-tagged events are shown in
Table~\ref{tab:r1-4-kaon} and Fig.~\ref{fig:r1-4-kaon-log}. 
For each
of the plots in this figure we calculate the Kolmogorov-Smirnov
probabilities for the PDF and data to originate from the same
distribution, as a way to verify the goodness of the fit. The
probabilities are 41\%, 99\%, 99\%, and 26\%.
Fig.~\ref{fig:asym} shows the raw, time-dependent $CP$ asymmetry 
\beq
A(\dt) = {N_{\stag=1}(\dt) - N_{\stag=-1}(\dt) 
		\over N_{\stag=1}(\dt) + N_{\stag=-1}(\dt)}.
\eeq
In the absence of background and with high statistics, perfect tagging, and 
perfect $\dt$ measurement, $A(\dt)$ would be a sinusoidal 
oscillation with amplitude $a_\dstpi$.

\begin{table}[!htb]
\begin{center}
\begin{tabular}{|c|c|c|}
\hline
 \multicolumn{2}{|c|}{Parameter} & Value                     \\ 
\hline
\multicolumn{3}{|c|}{signal} \\
  \hline
$2\r\sin(2\beta+\gamma)\cos\deltaPhase$ & $a_{\dstpi}^\ell$ &  $-0.048 \pm 0.022$  \\
$2\r\cos(2\beta+\gamma)\sin\deltaPhase$ & $c_{\dstpi}^\ell$ &  $-0.015 \pm 0.036$  \\
  \hline
$\Bz-\Bzb$ mixing frequency & $\dm{\dstpi} $ & $0.537 \pm 0.011$~ps$^{-1}$\\
$\Bz$ lifetime & $\lifetime{\dstpi} $ & $1.435 \pm 0.019$~ps\\
mistag (direct tags) & $\ot{\dstpi}^{\dir}$ &  $0.008 \pm 0.0.003$\\
\hline
bias of $\G^n_{cas}$ & $\bnj{\dstpi}{\cas}$ & $-0.43 \pm 0.22$\\
bias of $\G^n_{dir}$ & $\bnj{\dstpi}{\dir}$ & $0.000 \pm 0.034$\\
bias of $\G^w_{dir}$ & $\bwj{\dstpi}{\dir}$ & $0.04 \pm 0.40$\\
fraction of $\G^n_{dir}$ & $\fnj{\dstpi}{\dir}$ & $0.918 \pm 0.065$\\
fraction of $\G^o_{dir}$ & $\foj{\dstpi}{\dir}$ & $0.005 \pm 0.005$ \\
width of $\G^n_{dir}$ & $\snj{\dstpi}{\dir}$ & $1.047 \pm 0.063$\\
width of $\G^w_{dir}$ & $\swj{\dstpi}{\dir}$ & $2.43 \pm 0.68$\\
\hline
\multicolumn{3}{|c|}{continuum} \\
  \hline
$\qqbar$ effective lifetime & $\lifetime{\qqbar}$ & 1.12 $\pm$ 0.22~ps\\
mistag & $\ot{\qqbar} $ & 0.343 $\pm$ 0.011\\
mistag difference & $\Delta\ot{\qqbar} $ & $-0.095 \pm$ 0.022\\
fraction of 0-lifetime events & $f^\delta_{\qqbar} $ & 0.775 $\pm$ 0.037\\
\hline
bias of $\G^n$ & $\bn{\qqbar}$ & 0.047 $\pm$ 0.054 \\
bias of $\G^w$ & $\bw{\qqbar}$ & $-1.73 \pm$ 0.83 \\
fraction of $\G^n$ & $\fn{\qqbar}$ & 0.848 $\pm$ 0.055  \\
fraction of $\G^o$ & $\fo{\qqbar}$ & 0.072 $\pm$ 0.015 \\
width of $\G^n$ & $\sn{\qqbar}$ & 0.982 $\pm$ 0.052 \\
\hline
\end{tabular}
\caption{Results of the fit of the lepton-tagged events in the signal region
	$1.845 < \mmiss < 1.880$~\gevcc.
	Errors are statistical only.}
\label{tab:data_sr_leptag-Run1-4}
\end{center}
\end{table}

\begin{table}
\begin{center}
\begin{tabular}{|c|c|c|}  \hline
  \multicolumn{2}{|c|}{ Parameter} & Value                     \\ \hline
\multicolumn{3}{|c|}{signal} \\
  \hline
$2\r\sin(2\beta+\gamma)\cos\deltaPhase$ &  $a_\dstpi^K$                  & $-0.033 \pm 0.023$          \\
 $2\rp\sin(2\beta+\gamma)\cos\deltaPhaseP$ &  $b_\dstpi^K$                  & $-0.004 \pm 0.012$          \\
$2\cos(2\beta+\gamma)(\r\sin\deltaPhase - \rp\sin\deltaPhaseP)$ &   $c_\dstpi^K$                  & $0.019 \pm 0.023$          \\
\hline
$\Bz-\Bzb$ mixing frequency &   $\dm{\dstpi}$           & $0.4716\pm 0.0087$~ps$^{-1}$            \\
$\Bz$ effective lifetime &   $\lifetime{\dstpi}$     & $1.379\pm 0.012$~ps            \\
mistag &   $\ot{\dstpi}$           & $0.226\pm 0.0036$           \\
mistag difference &   $\dmistag{\dstpi}$      & $-0.0301\pm 0.0052$          \\
efficiency difference &   $\Delta \epsilon_{\dstpi}$      & $-0.0126\pm 0.0055$           \\
\hline
bias of $\G^n$ &   $\bn{\dstpi}$           & $-0.287\pm 0.022$             \\
bias of $\G^w$ &   $\bw{\dstpi}$           & $0.044\pm 0.067$             \\
fraction of $\G^n$ &   $\fn{\dstpi}$           & $0.909\pm 0.040$               \\
fraction of $\G^o$ &   $\fo{\dstpi}$           & $0.014\pm 0.002$              \\
width of $\G^n$ &   $\sn{\dstpi}$           & $1.109\pm 0.039$              \\ 
width of $\G^w$ &   $\sw{\dstpi}$           & $0.01\pm 0.23$              \\ 
\hline
\multicolumn{3}{|c|}{continuum} \\
  \hline
$\qqbar$ effective lifetime  &   $\lifetime{\qqbar}$  & $0.546\pm 0.037 $~ps             \\
mistag of non-0-lifetime events&   $\ot{\qqbar}^\tau$   & $0.000 \pm 0.002 $             \\
mistag of 0-lifetime events &   $\ot{\qqbar}^\delta$ & $0.334\pm 0.010 $             \\
fraction of 0-lifetime events &   $f^\delta_{\qqbar}$  & $0.768 \pm 0.021$             \\
\hline
bias of $\G^n$ & $\bn{\qqbar}$           &0.013 $\pm$  0.007    \\
bias of $\G^w$ &$\bw{\qqbar}$           &0.130$\pm$  0.047     \\
fraction of $\G^n$ &$\fn{\qqbar}$           &0.853 $\pm$  0.026     \\
fraction of $\G^o$ &$\fo{\qqbar}$           &0.013 $\pm$  0.001     \\
width of $\G^n$ &$\sn{\qqbar}$           &1.029 $\pm$  0.013     \\
width of $\G^w$ &$\sw{\qqbar}$           &1.90 $\pm$  0.11     \\ \hline
\hline
\end{tabular}
\caption{Results of the fit of the kaon-tagged events in the signal region
	$1.845 < \mmiss < 1.880$~\gevcc.
	Errors are statistical only.}
\label{tab:r1-4-kaon}
\end{center}
\end{table}

\clearpage

\section{SYSTEMATIC STUDIES}
\label{sec:Systematics}

The systematic errors are summarized in Tables~\ref{tab:syst-lepton}
and~\ref{tab:syst-kaon} for lepton- and kaon-tagged events, respectively.
Each item below corresponds to the item with the same number in
Tables~\ref{tab:syst-lepton} and~\ref{tab:syst-kaon}.

\begin{table}
\begin{center}
\begin{tabular}{|l|c|c|}  \hline
Source     & \multicolumn{2}{|c|}{Error $(\times 10^{-2})$ } \\
\hline 
		&  $ a_\dstpi^\ell$   
		&  $ c_\dstpi^\ell$    \\ 
\hline
1. Step~1 fit 			& 0.04 & 0.04   \\
2. Sideband statistics  	& 0.08 & 0.08   \\
3. $\fmis_\dstpi$        	& 0.02 & 0.02   \\
4. $\rho$          		& 0.02 & 0.02   \\ 
\hline
5. MC statistics    		& 0.6  & 1.2   \\
\hline
6. Beam spot size      		& 0.10	& 0.10   \\
7. Detector $z$ scale       	& 0.03	& 0.03   \\
8. Detector alignment 		& 0.4   & 0.8   \\
\hline
9. Combinatoric background   \CP content     & 0.25  & 0.22   \\
10. Peaking background \CP content    & 0.36  & 0.38   \\
11. $D^*\rho$    \CP content    & 0.53  & 0.52   \\ 
\hline
12. Peaking background   & 0.21    & 0.31   \\
13. Signal region/sideband difference     & 0.0003    &0.002    \\
14. \BR($\btodstrhopm$)  &  0.17   & 0.33   \\
15. Variation of  $\tau_\dstpi$ and $\Delta m_\dstpi$ & 0.21 & 0.95  \\
\hline
Total systematic error           & 1.04    & 1.89   \\ 
\hline
Statistical uncertainty     & 2.2 & 3.6 \\
\hline
\end{tabular}
\caption{Systematic errors in $a_\dstpi^\ell$ and $c_\dstpi^\ell$ for 
	lepton-tagged events. }
\label{tab:syst-lepton}
\end{center}
\end{table}

\begin{table}
\begin{center}
\begin{tabular}{|l|c|c|c|}  \hline
Source  & \multicolumn{3}{|c|}{Error $(\times 10^{-2})$ } \\
\hline
	        & $a_\dstpi^K$   
		&  $b_\dstpi^K$  
		& $c_\dstpi^K$  \\ \hline
1. Step~1 fit & $0.10 $& $0.04 $       & $0.04 $       \\
2. Sideband statistics &  $0.40  $   &   $0.12  $  &  $0.44  $   \\
3. $\fmis_\dstpi$        &  $0.02  $   &  negl.  $  $  &  negl.    \\
4. $\rho$          &  $0.02  $   &   negl.   &  negl.   \\
\hline
5. MC statistics    &  $0.8  $   &   $0.4  $  &  $0.9  $   \\ 
\hline
6. Beam spot size       &  $0.07  $   &   $0.13  $  &  $0.06  $   \\
7. Detector $z$ scale    &  $0.02  $   &   negl.               &  $0.03  $   \\
8. Detector alignment & $0.41  $ & $0.14  $    & $0.74  $   \\
\hline
9. Combinatoric background \CP content & $0.80$  & $0.56  $   &  $0.72  $   \\
10. Peaking background \CP content  &  $0.29$   &  $0.17  $   &  $0.27  $   \\
11. $D^*\rho$    \CP content    &  $0.57  $   &  $0.58  $   &  $0.58  $   \\ 
\hline
12. Peaking background    &  $0.21  $   &   $0.41  $  &  $0.31  $   \\
13. Signal region/sideband difference & $0.04$  &   $0.03$  &  $0.05$   \\
14. \BR($\btodstrhopm$)  &  $0.17  $   &   $0.22  $  &  $0.33  $   \\
15. Variation of $\tau_\dstpi$ and $\Delta m_\dstpi$ & $0.26$ & $0.16$ & $0.05$  \\
\hline
Total   systematic error        &  $1.47  $   &  $1.06  $   &  $1.63 $
\\ 
\hline
Statistical uncertainty     & 2.3 & 1.2 & 2.3 \\
\hline
\end{tabular}
\caption{Systematic errors in $a_\dstpi^K$, $b_\dstpi^K$, and $c_\dstpi^K$ for 
	kaon-tagged events.}
\label{tab:syst-kaon}
\end{center}
\end{table}

\begin{itemize}

\item[1.] The statistical errors from the fit in Step~1 are
propagated to the final fit, taking their correlations into
account. It also includes the systematic errors due to possible differences
between the PDF line shape and the data points in the kinematical fit.

\item[2.] The statistical errors from the $\mmiss$ 
sideband fit (Step~3) are propagated
to the final fit (Step~4), taking their correlations into account.

\item[3-4.] The statistical errors from the Step~2 
fits are propagated to the final fit.

\item[5.] The statistical errors associated with all the parameters
obtained from \mc\ are propagated to the final fit.

\item[6.] The effect of uncertainties in the 
beam-spot size on the vertex constraint 
is estimated by increasing the beam spot 
size by 50$~\mu$m.

\item[7.] The effect of the uncertainty
in the measured length of the detector in the $z$ direction
is evaluated by applying a 0.6\% variation to the measured values
of $\dt$ and $\dtErr$.

\item[8.] To evaluate the effect of possible misalignments in the
SVT, signal \mc\ events are reconstructed with different
alignment parameters, and the analysis is repeated.

\item[9-11.] The $CP$ parameters of the $\btodstrhopm$, peaking, and 
combinatoric
$\BB$ background are fixed to 0 in the fits. To study the effect of
possible \CP violation in these backgrounds, their \CP parameters are
varied in the range  $\pm 0.04$ and the step-4 fit is repeated.

\item[12.] The uncertainty due to the parameters of $\T'_\peak$
is evaluated by fitting the simulated sample, setting the 
parameters of $\T'_\peak$ to be identical to those of $\T'_\comb$.

\item[13.] The uncertainty due to possible differences between the
$\dt$ distributions for the combinatoric background in the $\mmiss$
sideband and signal region is evaluated by comparing the results of
fitting the simulated sample with the $\T'_\comb$ parameters taken
from the sideband or the signal region.

\item[14.] The ratio $f_\dstrho/f_\dstpi$ is varied by 
the uncertainty in the corresponding ratio of branching fractions,
obtained from Ref.~\cite{ref:pdg2004}.

\item[15.] The lifetime and mixing parameters from the fits are not very
consistent with the world average values when only the statistical
uncertainties from the fit are considered. However, the inconsistency is
significantly reduced when the estimated systematic uncertainties on the $\tau_\dstpi$ and
$\Delta m_\dstpi$ parameters are included. To determine the systematic errors
on the $\CP$ parameters due to $\tau_\dstpi$ and $\Delta m_\dstpi$, we repeat the fit
with these parameters fixed to their world-average values from
Ref.~\cite{ref:pdg2004}. Because of the small correlations between these parameters
and the $\CP$ parameters ($\leq 2\%$ for the kaon-tagged sample and $\leq
9\%$ for the lepton-tagged sample), the resulting changes in the $\CP$
parameters are small. These changes are assigned as systematic
uncertainties.
\end{itemize}

\section{PHYSICS RESULTS}
\label{sec:Physics}

Summarizing the values and uncertainties of the \CP parameters, we obtain
the following results from the lepton-tagged sample:
\bea
a_\dstpi^\ell   = 2 \r \sin(2\beta+\gamma)\cos\deltaPhase  &=& 
	-0.048 \pm 0.022 \pm 0.010, \nonumber\\
c_\dstpi^\ell   = 2 \r \cos(2\beta+\gamma)\sin\deltaPhase   &=& 
	-0.015 \pm 0.036 \pm 0.019. 
\label{eq:S-lepton}
\eea
The results from the kaon-tagged sample fits are
\begin{eqnarray}
a_\dstpi^K   = 2\r\sin(2\beta+\gamma)\cos\deltaPhase  &=& 
	-0.033 \pm 0.023 \pm 0.015, \nonumber\\
b_\dstpi^K   = 2\rp\sin(2\beta+\gamma)\cos\deltaPhaseP   &=& 
	-0.004 \pm 0.012 \pm 0.011, \nonumber\\
c_\dstpi^K = 2\cos(2\beta+\gamma)(\r\sin\deltaPhase - \rp\sin\deltaPhaseP)  &=&
	+0.019 \pm 0.023 \pm 0.016 .
\label{eq:abc-kaon}
\eea
Combining the results for lepton and kaon tags 
gives the amplitude of the time-dependent \CP asymmetry,
\begin{equation}
a_\dstpi= 2\r\sin(2\beta+\gamma)\cos\deltaPhase  = 
	-0.041 \pm 0.016~(stat.) \pm 0.010~(syst.),
\label{eq:combined-a}
\end{equation}
where the systematic error takes into account correlations between
the individual results.
This result deviates from zero by 2.2 standard deviations.


We use two methods for interpreting our results in terms of
constraints on $|\sinphi|$. Both methods involve minimizing a $\chi^2$
function that is symmetric under the exchange $\sinphi \rightarrow
-\sinphi$, and applying the method of Ref.~\cite{ref:Feldman}.
In the first method we make no assumption regarding the value of $\r$.
For different values of $\r$, we vary $\deltaPhase$ and $\sinphi$ so as to
minimize the function
\beq
\chi^2(\r, \sinphi, \deltaPhase) = \sum_{j,k=1}^3 \Delta x_j 
	V^{-1}_{jk} \Delta x_k , 
\eeq
where 
$\Delta x_j$ refers to the difference between the result of our measurement
of $a_\dstpi^K$, $a_\dstpi^\ell$, or $c_\dstpi^\ell$ (Eqs.~(\ref{eq:abc-kaon}) 
and~(\ref{eq:S-lepton})) and the
theoretical expressions given by Eq.~(\ref{eq:abc}).
The measurements of $b_\dstpi^K$ and $c_\dstpi^K$ are not used in the fit,
since they depend on the unknown values of $r'$ and $\delta'$.
The measurement error matrix $V$ is nearly diagonal, and
accounts for correlations between the measurements due to correlated
statistical and systematic uncertainties.
%
The parameters determined by this fit are 
$\sin (2 \beta + \gamma)$, which is limited to lie
in the range $[-1, 1]$, and $\deltaPhase$.
We then generate many parameterized 
MC experiments with the
same sensitivity as reported here for different values of $\sinphi$ 
and with $\deltaPhase=0$, which yields the 
most conservative lower limit on $|\sinphi|$. 
The fraction of these experiments in which
$\chi^2(\sinphi) - \chi^2_{\rm min}$ is smaller than in the data is
interpreted as the confidence level
(CL) of the limit on $|\sinphi|$.
The resulting 90\% CL lower limit on $|\sinphi|$ is shown 
as a function of $\r$ in
Fig.~\ref{fig:limit-vs-r}.
This limit is always based on the more conservative of the two possibilities
implied by the ambiguity $|\sinphi|\leftrightarrow |\cos\deltaPhase|$.

\begin{figure}[htb]
\begin{center}
        \includegraphics[width=0.7\textwidth]{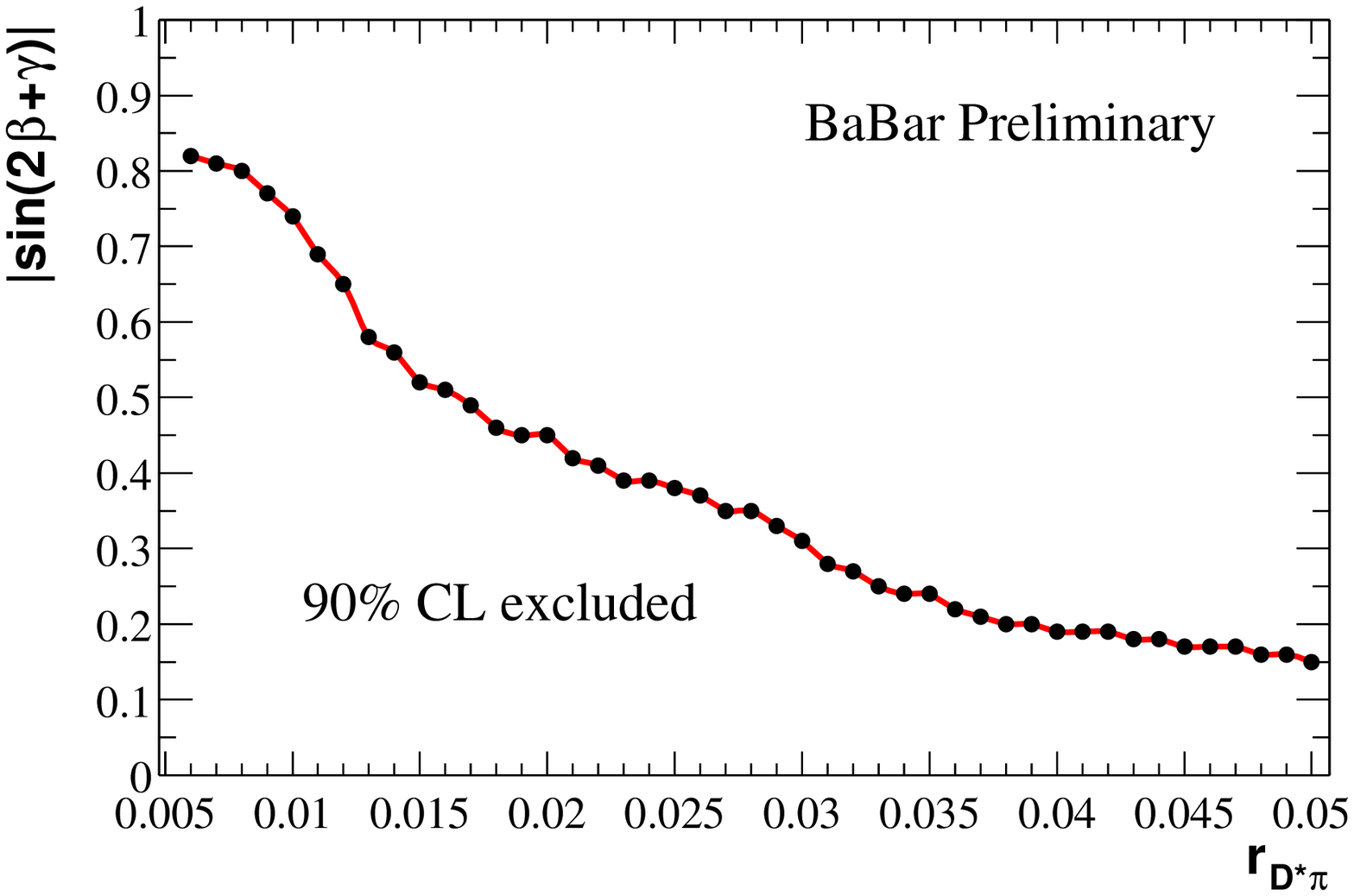}
\end{center}
\vspace*{-0.5cm}
\caption{
Lower limit on $|\sinphi|$ at 90\% CL as a function of $\r$. 
}
\label{fig:limit-vs-r}
\end{figure}


The second method assumes that $\r$ can be estimated from the
Cabibbo angle, the ratio of branching fractions ${\cal
B}(B^0\rightarrow {\Dstar}_s^{+} \pi^-) / {\cal B}(B^0\rightarrow
{\Dstar}^{-} \pi^+)$~\cite{ref:Dspi}, and the ratio of decay constants
$f_{\Dstar} / f_{\Dstar_s}$~\cite{ref:dec-const}, 
yielding the measured value
\beq
\rmeas = 0.015^{+0.004}_{-0.006}.
\label{eq:rmeas}
\eeq
This value includes our recent measurement of the branching fraction
$\B(D_s^+\to \phi\pi^+)$~\cite{ref:phipi}.
In addition to the above experimental errors, we attribute
a non-Gaussian 30\% relative error to the theoretical assumptions
involved in obtaining this value. 
To obtain the limits with these assumptions, we minimize the function
\beq
\tilde{\chi}^2 = \chi^2 + \Delta^2(\r),
\eeq
where the term $\Delta^2(\r)$ takes into account both the
Gaussian experimental errors of Eq.~(\ref{eq:rmeas}) 
and the 30\% theoretical uncertainty~\cite{Hocker:2001xe}: 
\beq
\Delta^2(\r) = \left\{\matrix{ 
	\left(\dfrac{\r - 1.3\, \rmeas }{ 0.005}\right)^2  & , &       
		\xi_\r          > 0.3 &, \cr
	0 & , & \left|\xi_\r\right| \le 0.3 &,\cr
	\left(\dfrac{\r - 0.7\, \rmeas }{ 0.007}\right)^2 & , &       
		\xi_\r          < -0.3 &,
}\right.  
\eeq
where $\xi_\r \equiv {(\r - \rmeas) / \rmeas}$.
The parameters 
\bea
\sin (2 \beta + \gamma) &=& 0.97 \pm 1.22, \nonumber\\
\deltaPhase &=& 0.24   \pm 1.23, \nonumber\\
\r &=& 0.020 \pm 0.004
\label{eq:fitresult}
\eea
are determined in this fit.
Due to the fact that the minimum value occurs close to the boundary of the 
physical region ($|\sin (2 \beta + \gamma)|=1$), the errors  
in Eq.~\ref{eq:fitresult} are not relevant and in order to give a 
probabilistic interpretation the method of Ref.~\cite{ref:Feldman} has  been used. 
The resulting confidence level as a function of the lower limit on
$|\sinphi|$, evaluated using parameterized MC experiments
as in the first method, is shown in Fig.~\ref{fig:FC1}. 
In particular, we find the limits
$|\sinphi|> 0.75~(0.58)$ at 68\% (90\%) CL.
The implied contours of constant probability for the apex of the
unitary triangle appear in Fig.~\ref{fig:UT}.

\begin{figure}[htb]
\begin{center}
        \includegraphics[width=0.7\textwidth]{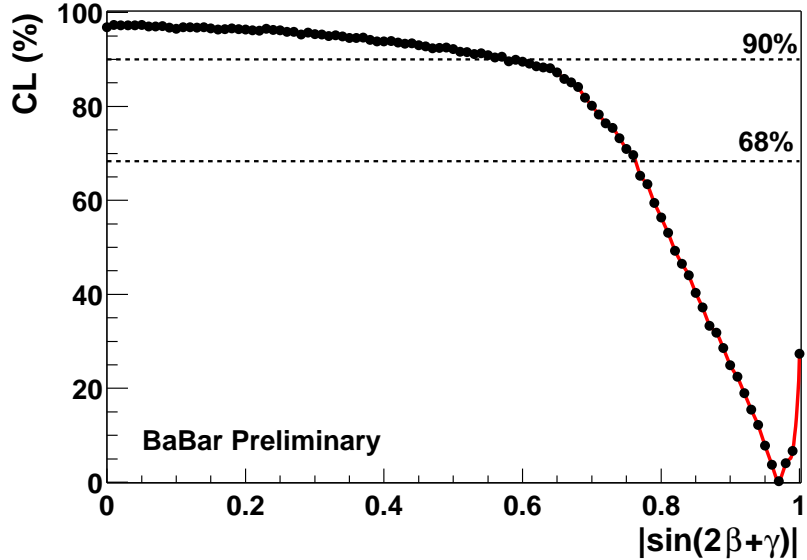}
\end{center}
\vspace*{-0.5cm}
\caption{
Confidence level as a function of the lower limit on $|\sinphi|$
given $\r$ from Eq.~(\ref{eq:rmeas}).
}
\label{fig:FC1}
\end{figure}

\begin{figure}[htb]
\begin{center}
        \includegraphics[width=0.7\textwidth]{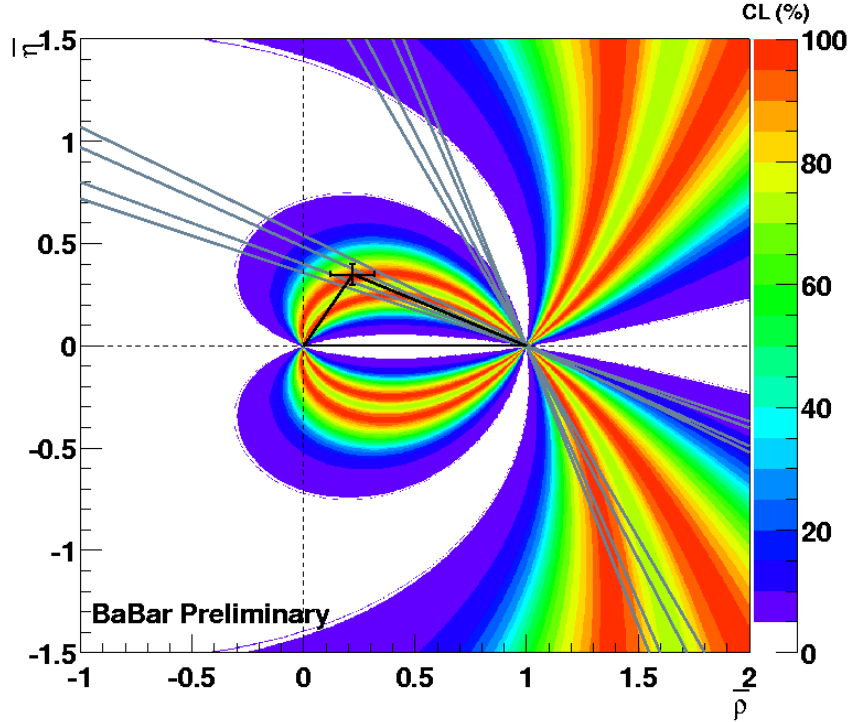}
\end{center}
\vspace*{-0.5cm}
\caption{
Contours of constant probability (color-coded in percent) 
for the position of the apex of the unitary triangle
to be inside the contour, based on the results of Fig.~\ref{fig:FC1}. 
The one- and two-standard deviation ranges of the world average $\sin2\beta$
measurement ($\sin2\beta=0.736\pm0.049$) are shown as gray lines. The 
cross represents the value
and errors of the apex of the unitarity triangle from the standard
CKMFitter fit \cite{ref:apex}.
}
\label{fig:UT}
\end{figure}

\section{SUMMARY}
\label{sec:Summary}

We present a preliminary measurement of the time-dependent \CP asymmetry and
parameters related to $\sinphi$ in a sample of partially reconstructed
$\btodstpi$ events.  In particular, the amplitude of the measured
asymmetry is
\begin{equation}
a_\dstpi= 2\r\sin(2\beta+\gamma)\cos\deltaPhase  = 
	-0.041 \pm 0.016~(stat.) \pm 0.010~(syst.).
\end{equation}
We interpret our results in terms of the lower limits
$|\sinphi|> 0.75~(0.58)$ at 68\% (90\%) CL, 
and extract limits as a function of the ratio $\r$
between the $b\rightarrow u \overline c
d$ and $b\rightarrow c \overline u d$ decay amplitudes.

\section{ACKNOWLEDGMENTS}
\label{sec:Acknowledgments}

We are grateful for the 
extraordinary contributions of our \pep2\ colleagues in
achieving the excellent luminosity and machine conditions
that have made this work possible.
The success of this project also relies critically on the 
expertise and dedication of the computing organizations that 
support \babar.
The collaborating institutions wish to thank 
SLAC for its support and the kind hospitality extended to them. 
This work is supported by the
US Department of Energy
and National Science Foundation, the
Natural Sciences and Engineering Research Council (Canada),
Institute of High Energy Physics (China), the
Commissariat \`a l'Energie Atomique and
Institut National de Physique Nucl\'eaire et de Physique des Particules
(France), the
Bundesministerium f\"ur Bildung und Forschung and
Deutsche Forschungsgemeinschaft
(Germany), the
Istituto Nazionale di Fisica Nucleare (Italy),
the Foundation for Fundamental Research on Matter (The Netherlands),
the Research Council of Norway, the
Ministry of Science and Technology of the Russian Federation, and the
Particle Physics and Astronomy Research Council (United Kingdom). 
Individuals have received support from 
CONACyT (Mexico),
the A. P. Sloan Foundation, 
the Research Corporation,
and the Alexander von Humboldt Foundation.

\end{document}